\documentclass[aps,prb,twocolumn,superscriptaddress,showpacs]{revtex4-2}
\usepackage{amssymb}
\usepackage{graphicx}
\usepackage{amsmath}
\usepackage{color}
 \usepackage{multirow}
\usepackage{booktabs}
\usepackage{calc}
\usepackage[colorlinks,linkcolor=blue, urlcolor=blue, anchorcolor=black, citecolor=blue, linkcolor=black]{hyperref}

\begin{document}

\title[]{ConfPred: A layered intergrowth structure prediction model based on confinement self-assembly in two-dimensional interlayer space}
\author{Hao Jiang} 
\email[E-mail address:]{hjiang@xtu.edu.cn}
\affiliation{Department of Physics, Zhejiang University, Hangzhou 310027, China}
\affiliation{School of Physics and Optoelectronics, Xiangtan University, Xiangtan 411105, China}
%\email[E-mail address:]{jialay@zju.edu.cn}
\author{HuiXiang Chen}
\affiliation{School of Physics and Optoelectronics, Xiangtan University, Xiangtan 411105, China}
\author{GuangHan Cao}
\email[E-mail address:]{ghcao@zju.edu.cn}
\affiliation{Department of Physics, Zhejiang University, Hangzhou 310027, China}
\begin{abstract}
We constructed a simple but effective model to predict the layered intergrowth structures by combining the self-assembly phenomenon in confined space and the sandwich configuration of layered materials. In this model, a two-dimensional confined space is constructed by two known block layers, such as the Fe$_2$As$_2$ block of iron-based superconductors. Then, the crystal structure prediction is carried out only inside the confined space to search for brand-new block layers. We realized this model on the basis of the USPEX9.4 code. In the test, the already existing iron-based superconductors can be always successfully found, such as Ba$_2$Ti$_2$Fe$_2$As$_4$O, Sr$_3$Sc$_2$Fe$_2$As$_2$O$_5$, Sr$_4$V$_2$Fe$_2$As$_2$O$_6$, and so on. The comparison test suggests that our model has remarkable advantages in searching for intergrowth structures. With this space confinement prediction model, a structure prediction of layered intergrowth materials even with up to six elements can be performed with an acceptable machine time consumption. So far, we have done some multi-composition crystal structure predictions of iron-based superconductor, and found several stable and metastable structures, such as Ba$_2$Fe$_2$As$_3$, Eu$_2$Fe$_2$As$_3$, La$_2$O$_2$ClFeAs, LiOMn$_2$As,Li$_4$OFe$_2$As$_2$. 
\end{abstract}
\maketitle
It is increasingly popular for the computational approaches to be applied in finding superconductors, such as crystal structure prediction (CSP), machine learning, high-throughput prediction, and their combinations. One of the most remarkable works is the successful prediction of the LaH$_{10}$ high-pressure superconductor, whose superconducting critical temperature under high pressure is as high as 250 K \cite{10.1073/pnas.1704505114,10.1038/s41586-019-1201-8,10.1103/physrevlett.122.027001}. It stimulated a blowout in the structural prediction of high-pressure superconductors \cite{10.1103/physrevlett.123.097001,Salke.2020,Tsuppayakorn-aek.2021,Shi.2021,Liang.2021}. The current efforts of predicting superconductors are mainly concentrated in conventional superconductors. Generally, most unconventional superconductors are systems with more than three elements. The Variable-composition CSP for the quaternary systems is too challenging to reach \cite{10.1039/c8fd90033g}. Machine learning and high-throughput method are also racing to be applied to search for new superconductors \cite{10.7567/1882-0786/ab2922,Hutcheon.2020,Yang.2021,10.1088/1361-6668/ab51b1,Yu.2020}. However, their applications in unconventional HTSCs are still rare because of the remaining challenges. It is in great demand to develop a reliable and efficient method to searching for HTSCs comprising more than three elements \cite{10.7567/1882-0786/ab2922}.

The CSP can be divided into two aspects of problems \cite{10.1039/c8fd90033g}. One is the problem of ranking, that is, the reliable calculation of the lattice enthalpy. The strong correlation and the $Van$ $de$ $Waals$ force between the layers of unconventional HTSCs affect the reliability of the first-principles calculation to some extent. The second is the search problem, which is to find out the possible arrangements of atoms in the crystal lattice. Usually, the size of the configuration space increases exponentially with the number of atoms\cite{10.1016/j.physrep.2020.02.003}. The structure of unconventional HTSCs is more complicated and contains more elements \cite{Jiang.2013i1j}. Much of unconventional HTSCs contain up to 4-6 elements, making the direct application of the current CSP approaches too expensive computationally. 

The structure of unconventional HTSCs is relatively complex and diverse but with a simple characteristic. It is always the specific block layer that is responsible for the superconductivity. Such as the CuO plain in Cu-based superconductors and Fe$_2X_2$ (“$X$” refers to As, P, Se, or Te) block layer in iron-based superconductors. The interlayer space of the superconducting layers is usually filled by cation, negative block layer, positive block layer, or their combination, forming the sandwich configuration \cite{Jiang.2013i1j,Gui.2021}. Generally, the superconducting layer and other block layers are relatively stable in the sandwich structure. Sometimes, the design of new superconductors is considered as building Lego blocks \cite{10.1088/0034-4885/79/7/074502}. Suppose the lattice parameters of blocks, the charge balance, the rule of ``Hard and Soft Acids and Bases" \cite{Pearson.1963}, and so on, can meet the corresponding requirements \cite{Jiang.2013i1j}. In that case, it is possible to form a final stable superconducting structure. Inspired by the sandwich configuration, we propose a new structure prediction model based on the confinement self-assembly, which greatly reduces the difficulty of the search problem and is also beneficial to the ranking problem. We first construct an appropriate two-dimensional confined space with two known stable layers and then let the CSP be carried out inside the interlayer confined space. Namely, all the operations of CSP are only valid for ions inside the two-dimensional confined space, and the DFT relaxation is performed on the whole structure. The confinement self-assembly phenomenon has been widely used in nanotubes, catalysis, two-dimensional carbon nano-films, and so on \cite{10.1002/anie.201712959,Shamay.2018,10.1021/acs.chemmater.7b03349,10.1016/j.jmat.2017.03.001,Frederix.2015}. Its application in layered structure prediction is worth looking forward to.

\begin{figure*}
% \flushright
\centering

%\begin{center}
\includegraphics [width=6.5in]{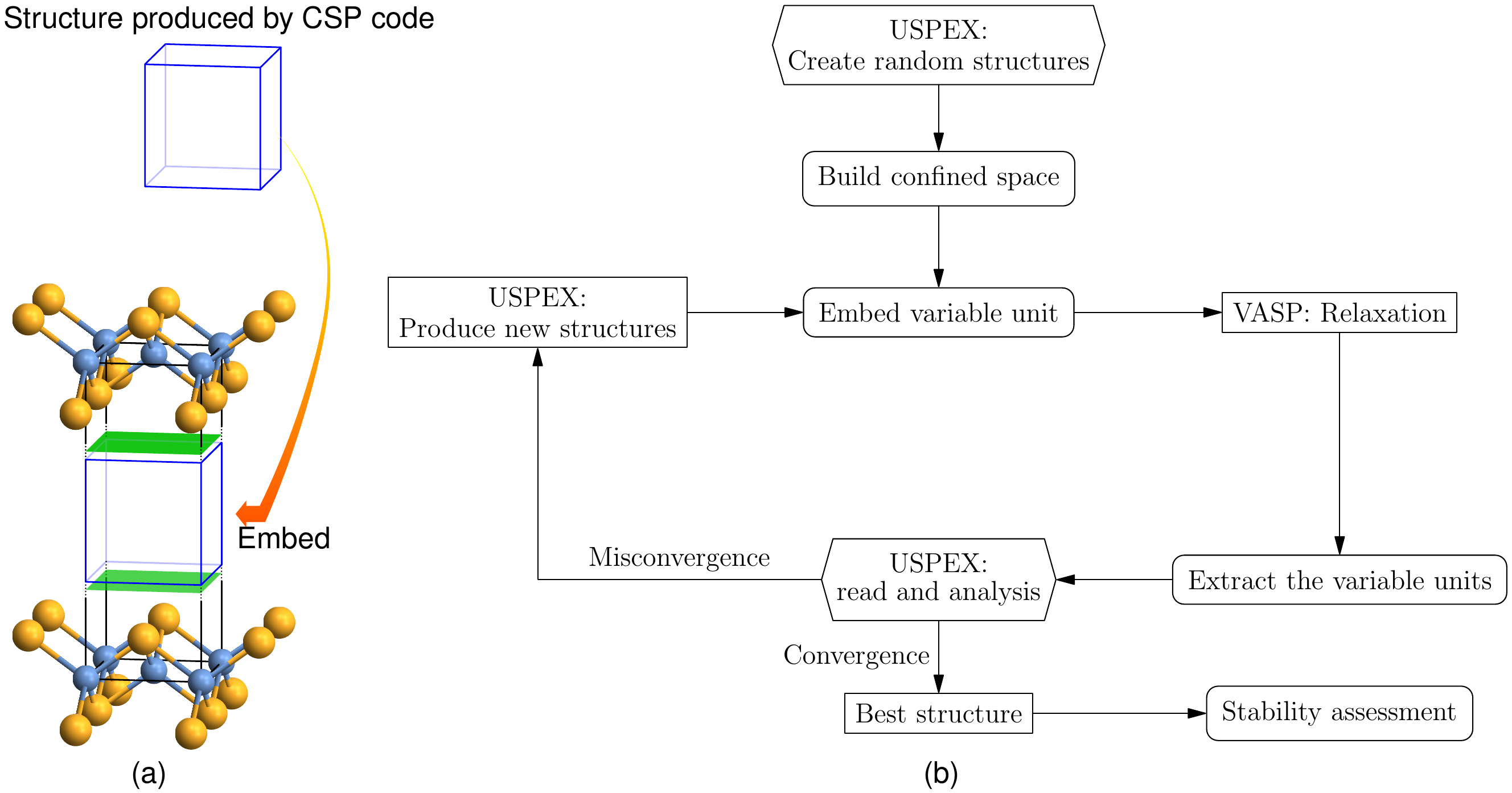}
\caption{(a) The antifluorite-like block layers are the fixed units. The space between the blue planes is the two-dimensional confined space, whose size will be relaxed by the VASP code. The blue box denotes the structure produced by the CSP programs. (b) The flow chart of CSP with this model. The rectangle boxes indicate the functions of the original CSP code. The round rectangle boxes represent the functions we added. The functions in the bevel boxes are modified accordingly. The one iteration of Steps 3$\rightarrow$4$\rightarrow$5$\rightarrow$6$\rightarrow$7$\rightarrow$3 is the process of one generation.}\label{f1}
%\end{center}
\end{figure*}

\section{Model and implementation}
The block layers in the unconventional HTSCs have a certain degree of independence, and the structural design of HTSCs is much like building Lego blocks. Here, we treat the invariable superconducting layer as the fixed structural unit or fixed unit for short. The staffs (ions and other layers) between the superconducting layers are considered as the variable structural unit or variable unit for short. Briefly, we construct the model as follows: Build a two-dimensional confined space with two fixed units, as shown in Fig. \ref{f1} (a), and then perform CSP within this confined space. All the operations by CSP are only valid for the variable unit inside the confined space. As this form of CSP cannot be realized with the current CSP programs directly, we carried out the CSP of the above model by modifying the source code of USPEX9.4 \cite{Glass.2006,10.1021/ar1001318,Pakhnova.2020}. Here we take iron-based superconductors as an example to describe the details of this model. The implementation follows: Step 1, select some appropriate ions and generate random structures through the CSP program, such as USPEX (A reliable CSP program that has been widely used \cite{10.1038/s41467-021-23677-w,Liu.2021}.). Now, the random structures are the variable structural units. Step 2, build a suitable two-dimensional confined space with two Fe$_2$As$_2$ layers in the tetragonal lattice. Step 3, embed the variable unit into the confined space as demonstrated in Fig. \ref{f1}. Step 4, relax the combinational structures with the VASP code \cite{Kresse.1993}. Step 5, after the relaxation, extract the variable unit from the combinational structures for further analysis by the CSP program. Step 6, Read and analyze the calculation results in the CSP program. Step 7, Produce a new generation of structures in the CSP program. Then, embed these structures into the confined space again for the subsequent relaxation, and iterate the above process until it reaches convergence. The flow of the above implementation process is listed in Fig. \ref{f1} (b).
\newcommand{\tabincell}[2]{\begin{tabular}{@{}#1@{}}#2\end{tabular}}  
\begin{table*}
\caption{Comparative test. All of the seeds we applied include the Fe$_2$As$_2$ block layer, and the ions structure of the other part is generated randomly, such as the seed structure 18 shown in Fig. \ref{f3}(a). }
\begin{tabular}{c|ccc|ccc}
\hline
\hline
Target structure  & \multicolumn{3}{c|}{SrFFeAs} & \multicolumn{3}{c}{Ba$_2$Ti$_2$Fe$_2$As$_4$O} \\
 \hline
\multirow{2}{*}{Sides to compare} & \multicolumn{2}{c}{USPEX} & \quad ConfPred \qquad & \multicolumn{2}{c}{USPEX} & \quad ConfPred \qquad \\
          & \quad Without Seed \qquad &\quad With Seed \qquad  & \quad  \qquad & \quad Without Seed \qquad  & \quad With Seed\qquad & \\
\hline
Evolution process &   Fig. \ref{f5} (a)    &  Fig. \ref{f2} (a)    &   Fig. \ref{f2} (b)    &    Fig. \ref{f5} (b)   &   Fig. \ref{f3} (a)    &  Fig. \ref{f3} (b)    \\
Final best structure& \quad 1111 phase \quad &\quad 1111 phase \quad &\quad 1111 phase\quad & \quad Fig. 5(b)-77 \quad & \quad Fig. 3(a)-119 \quad & \quad 22241 phase \quad\\
Total enthalpy (eV/formula) &  -23.62 & -23.62  & -23.62 & -69.09& -69.50 & -71.38  \\
Space group  &  129     &   129   &  129  &   1    &   44   & 139       \\
Total generations &  11     &    8   &    5   &   11    &14       &  13   \\
Population &   111    &  86    &   56    &    128   &   172    &  162    \\
 \hline
 \hline
\end{tabular}
\label{t1}
\end{table*}

Further explanation for some aspects of the above process are detailed in the following:
\newline
A) The stacking pattern of the two fixed units that are used to build the confined space is usually not unique. Such as that Fe$_2$As$_2$ superconducting layer and similar block layers has two different stacking patterns generally. One is that the upper and lower superconducting layers are stacked in translational symmetry, as shown in Fig. \ref{f1}(a). The other is stacked with one of the block layer shifted by half lattice constant along the crystallographic axis a or b, which will form a lattice with inversion symmetry. Most of the time the fixed units stacked in translational symmetry can also find a stable structure in inversion symmetry with the form of a Niggli reduced cell \cite{10.1107/s0567739476000636} or a primitive cell. While, It would be more powerful to give the possible stacking patterns in the beginning in searching for complex structures, such as the instance shown in Fig. \ref{f3} (b).
\newline
B) The embedding patterns of the variable structure unit could be many. The translation, rotation, and their combination of the variable structure unit would result in entirely different combinational structures. We traverse all possible operations to generate all of the possible combinational structures. This process will produce some repetitive structures. We delete these repetitive structures through the structure fingerprint. Then the static enthalpy of the remaining structures is calculated with VASP. The structure with the lowest enthalpy will be pushed to the next step of the relaxation process. Because these structures are only embedded in different ways, their total enthalpy only needs to be calculated with a very low accuracy so that this step can be done quickly. 
\newline
C) Generally, the distance of ions in CSP programs is restricted to improve efficiency. According to the above model, the restricted condition can be modified appropriately to improve efficiency further. The distance between the top and bottom ions of the variable unit should not be restricted because the translational symmetry along the c-direction of the crystallographic axis will be broken after the embedding action. Besides, a simple but efficient restricted condition can be added to the process of generating random structures. Such as for the search of the iron-based superconductors, the ions at the upper and lower sides of the variable unit can be restricted as cations because the anion will repel another anion As$^{3-}$ strongly, which will significantly increase the energy of the whole structure.

The design of a structure with specific functions is an art. This model greatly simplifies the design of intergrowth materials, but picking the suitable elements from the periodic table to build the variable structure unit is still a highly skilled task. Here, we summarize some rules for reference:
\newline
i) The rule of ``Hard and Soft Acids and Bases" \cite{Pearson.1963}. Generally for iron-based superconductors, hard acids are more likely to form stable bonds with hard bases, while soft acids are more likely to bond with soft bases \cite{Jiang.2013i1j,Zhi-Cheng.2018,Wang.2021ca}. Such as LaOFeAs, the hard acid La$^{3+}$ combines with the hard base O$^{2-}$, meanwhile the soft acid Fe$^{2+}$ bonds with the soft base As$^{3-}$.
\newline
ii) Try to avoid mixed occupations. The selected elements should have few chance to mix with the fixed structure unit \cite{Jiang.2013i1j} . The mixed occupations will introduce disorder, which increases the enthalpy of the structure.
\newline
iii) Keep the stability of the fixed unit. For example, The hard base should be neutralized with a hard acid to avoid REDOX reaction with the fixed unit \cite{Jiang.2013i1j}.
\newline
iv) Try to meet the charge balance. The valence state of the fixed unit is adjustable by the variable unit, but try not to exceed its adaptable range \cite{Jiang.2013i1j}. The charge balance is an important factor for considering the number of atoms adopted in the variable unit. 

% \section{Stability assessment}
The convex hull construction is a reliable tool to determine the stability of a compound. Usually, a convex hull diagram of a compound with 2-3 elements is not too difficult to draw, but the situation of compounds with up to four elements will change because of too many dimensions to illustrate. The line in the convex hull diagram means linear superposition. We wrote a program automatically to calculate the possible superpositions as more as possible. With this tool, we can find the most stable decomposer of a target compound. As a result, we find that it is a highly effective way to determine the stability of a compound. The test in the 1111 phase of iron-based superconductors is listed in table \ref{tn1}. Their formation enthalpy well denotes the stabilities.

\begin{table*}[]
\caption{Based on the formation enthalpy, the stability of the La-based 1111 phase in iron superconductors are assessed. The data of total enthalpy are supported by the ``Material project'' or calculated by Custodian + VASP\cite{10.1016/j.commatsci.2017.07.030,10.1016/j.commatsci.2012.10.028}. The spin of magnetic atoms are considered with Custodian}
\begin{tabular}{c|cc|c|c}
\hline
\hline
Compound \&      & Decomposers \&    &   &  E$_\text{form}$ & Synthesis\\
energy (eV/formula) & energy (eV/formula)& & (eV/formula) & \\
\hline 
$\mathrm{La}\mathrm{O}\mathrm{Fe}\mathrm{As}$ & $\hspace{5.500 pt}\frac{1}{3}\mathrm{Fe}\hspace{5.500 pt}$+$\hspace{5.500 pt}\frac{1}{3}\mathrm{La}_{2}\mathrm{O}_{3}\hspace{5.500 pt}$+$\hspace{5.500 pt}\frac{1}{3}\mathrm{La}\mathrm{As}\hspace{5.500 pt}$+$\hspace{5.500 pt}\frac{2}{3}\mathrm{Fe}\mathrm{As}\hspace{5.500 pt}$ & Total & & \multirow{2}{*}{yes}\\
-30.32 & \hspace{3.207 pt}-2.82\hspace{10.985 pt}\hspace{9.638 pt}-13.97\hspace{17.415 pt}\hspace{9.485 pt}-4.24\hspace{17.263 pt}\hspace{8.929 pt}-9.02\hspace{8.929 pt} & -30.06 & -0.26 & \\
\hline
$\mathrm{Ce}\mathrm{O}\mathrm{Fe}\mathrm{As}$ & $\hspace{5.500 pt}\frac{5}{12}\mathrm{Fe}\hspace{5.500 pt}$+$\hspace{5.500 pt}\frac{1}{12}\mathrm{Ce}_{7}\mathrm{O}_{12}\hspace{5.500 pt}$+$\hspace{5.500 pt}\frac{5}{12}\mathrm{Ce}\mathrm{As}\hspace{5.500 pt}$+$\hspace{5.500 pt}\frac{7}{12}\mathrm{Fe}\mathrm{As}\hspace{5.500 pt}$ & Total & & \multirow{2}{*}{yes}\\
-31.27 & \hspace{5.200 pt}-3.53\hspace{12.978 pt}\hspace{13.832 pt}-13.93\hspace{21.610 pt}\hspace{11.686 pt}-5.61\hspace{19.464 pt}\hspace{10.922 pt}-7.90\hspace{10.922 pt} & -30.96 & -0.31 & \\
\hline
$\mathrm{Pr}\mathrm{O}\mathrm{Fe}\mathrm{As}$ & $\hspace{5.500 pt}\frac{1}{3}\mathrm{Fe}\hspace{5.500 pt}$+$\hspace{5.500 pt}\frac{1}{3}\mathrm{Pr}_{2}\mathrm{O}_{3}\hspace{5.500 pt}$+$\hspace{5.500 pt}\frac{1}{3}\mathrm{Pr}\mathrm{As}\hspace{5.500 pt}$+$\hspace{5.500 pt}\frac{2}{3}\mathrm{Fe}\mathrm{As}\hspace{5.500 pt}$ & Total & & \multirow{2}{*}{yes}\\
-29.76 & \hspace{3.207 pt}-2.82\hspace{10.985 pt}\hspace{9.374 pt}-13.67\hspace{17.151 pt}\hspace{9.221 pt}-4.14\hspace{16.999 pt}\hspace{8.929 pt}-9.02\hspace{8.929 pt} & -29.66 & -0.09 & \\
\hline
$\mathrm{Nd}\mathrm{O}\mathrm{Fe}\mathrm{As}$ & $\hspace{5.500 pt}\frac{1}{3}\mathrm{Fe}\hspace{5.500 pt}$+$\hspace{5.500 pt}\frac{1}{3}\mathrm{Nd}_{2}\mathrm{O}_{3}\hspace{5.500 pt}$+$\hspace{5.500 pt}\frac{1}{3}\mathrm{Nd}\mathrm{As}\hspace{5.500 pt}$+$\hspace{5.500 pt}\frac{2}{3}\mathrm{Fe}\mathrm{As}\hspace{5.500 pt}$ & Total & & \multirow{2}{*}{yes}\\
-29.82 & \hspace{3.207 pt}-2.82\hspace{10.985 pt}\hspace{10.540 pt}-13.74\hspace{18.318 pt}\hspace{10.387 pt}-4.15\hspace{18.165 pt}\hspace{8.929 pt}-9.02\hspace{8.929 pt} & -29.74 & -0.08 & \\
\hline
$\mathrm{Pm}\mathrm{O}\mathrm{Fe}\mathrm{As}$ & $\hspace{5.500 pt}\mathrm{Fe}\hspace{5.500 pt}$+$\hspace{5.500 pt}\frac{1}{3}\mathrm{Pm}_{2}\mathrm{O}_{3}\hspace{5.500 pt}$+$\hspace{5.500 pt}\frac{1}{3}\mathrm{Pm}\mathrm{As}_{3}\hspace{5.500 pt}$ & Total & & \multirow{2}{*}{yes}\\
-29.90 & \hspace{0.014 pt}-8.47\hspace{7.792 pt}\hspace{11.582 pt}-13.83\hspace{19.360 pt}\hspace{13.672 pt}-6.93\hspace{13.672 pt} & -29.23 & -0.66 & \\
\hline
$\mathrm{Sm}\mathrm{O}\mathrm{Fe}\mathrm{As}$ & $\hspace{5.500 pt}\frac{1}{3}\mathrm{Fe}\hspace{5.500 pt}$+$\hspace{5.500 pt}\frac{1}{3}\mathrm{Sm}_{2}\mathrm{O}_{3}\hspace{5.500 pt}$+$\hspace{5.500 pt}\frac{1}{3}\mathrm{Sm}\mathrm{As}\hspace{5.500 pt}$+$\hspace{5.500 pt}\frac{2}{3}\mathrm{Fe}\mathrm{As}\hspace{5.500 pt}$ & Total & & \multirow{2}{*}{yes}\\
-29.89 & \hspace{3.207 pt}-2.82\hspace{10.985 pt}\hspace{10.957 pt}-13.84\hspace{18.735 pt}\hspace{10.804 pt}-4.14\hspace{18.582 pt}\hspace{8.929 pt}-9.02\hspace{8.929 pt} & -29.83 & -0.06 & \\
\hline
$\mathrm{Gd}\mathrm{O}\mathrm{Fe}\mathrm{As}$ & $\hspace{5.500 pt}\frac{1}{3}\mathrm{Fe}\hspace{5.500 pt}$+$\hspace{5.500 pt}\frac{1}{3}\mathrm{Gd}_{2}\mathrm{O}_{3}\hspace{5.500 pt}$+$\hspace{5.500 pt}\frac{1}{3}\mathrm{Gd}\mathrm{As}\hspace{5.500 pt}$+$\hspace{5.500 pt}\frac{2}{3}\mathrm{Fe}\mathrm{As}\hspace{5.500 pt}$ & Total & & \multirow{2}{*}{yes}\\
-39.30 & \hspace{3.207 pt}-2.82\hspace{10.985 pt}\hspace{10.714 pt}-20.14\hspace{18.492 pt}\hspace{10.561 pt}-7.27\hspace{18.339 pt}\hspace{8.929 pt}-9.02\hspace{8.929 pt} & -39.26 & -0.04 & \\
\hline
$\mathrm{Tb}\mathrm{O}\mathrm{Fe}\mathrm{As}$ & $\hspace{5.500 pt}\frac{1}{3}\mathrm{Fe}\hspace{5.500 pt}$+$\hspace{5.500 pt}\frac{1}{3}\mathrm{Tb}_{2}\mathrm{O}_{3}\hspace{5.500 pt}$+$\hspace{5.500 pt}\frac{1}{3}\mathrm{Tb}\mathrm{As}\hspace{5.500 pt}$+$\hspace{5.500 pt}\frac{2}{3}\mathrm{Fe}\mathrm{As}\hspace{5.500 pt}$ & Total & & \multirow{2}{*}{High pressure}\\
-29.97 & \hspace{3.207 pt}-2.82\hspace{10.985 pt}\hspace{10.401 pt}-14.00\hspace{18.179 pt}\hspace{10.249 pt}-4.12\hspace{18.026 pt}\hspace{8.929 pt}-9.02\hspace{8.929 pt} & -29.97 & -0.01 & \\
\hline
$\mathrm{Dy}\mathrm{O}\mathrm{Fe}\mathrm{As}$ & $\hspace{5.500 pt}\frac{1}{3}\mathrm{Fe}\hspace{5.500 pt}$+$\hspace{5.500 pt}\frac{1}{3}\mathrm{Dy}_{2}\mathrm{O}_{3}\hspace{5.500 pt}$+$\hspace{5.500 pt}\frac{1}{3}\mathrm{Dy}\mathrm{As}\hspace{5.500 pt}$+$\hspace{5.500 pt}\frac{2}{3}\mathrm{Fe}\mathrm{As}\hspace{5.500 pt}$ & Total & & \multirow{2}{*}{High pressure}\\
-29.97 & \hspace{3.207 pt}-2.82\hspace{10.985 pt}\hspace{10.540 pt}-14.01\hspace{18.318 pt}\hspace{10.387 pt}-4.11\hspace{18.165 pt}\hspace{8.929 pt}-9.02\hspace{8.929 pt} & -29.97 & 0.00 & \\
\hline
$\mathrm{Ho}\mathrm{O}\mathrm{Fe}\mathrm{As}$ & $\hspace{5.500 pt}\frac{1}{3}\mathrm{Fe}\hspace{5.500 pt}$+$\hspace{5.500 pt}\frac{1}{3}\mathrm{Ho}_{2}\mathrm{O}_{3}\hspace{5.500 pt}$+$\hspace{5.500 pt}\frac{1}{3}\mathrm{Ho}\mathrm{As}\hspace{5.500 pt}$+$\hspace{5.500 pt}\frac{2}{3}\mathrm{Fe}\mathrm{As}\hspace{5.500 pt}$ & Total & &\multirow{2}{*}{High pressure} \\
-29.95 & \hspace{3.207 pt}-2.82\hspace{10.985 pt}\hspace{10.263 pt}-14.04\hspace{18.040 pt}\hspace{10.110 pt}-4.09\hspace{17.888 pt}\hspace{8.929 pt}-9.02\hspace{8.929 pt} & -29.98 & 0.03 & \\
\hline
$\mathrm{Er}\mathrm{O}\mathrm{Fe}\mathrm{As}$ & $\hspace{5.500 pt}\frac{1}{3}\mathrm{Fe}\hspace{5.500 pt}$+$\hspace{5.500 pt}\frac{1}{3}\mathrm{Er}_{2}\mathrm{O}_{3}\hspace{5.500 pt}$+$\hspace{5.500 pt}\frac{1}{3}\mathrm{Er}\mathrm{As}\hspace{5.500 pt}$+$\hspace{5.500 pt}\frac{2}{3}\mathrm{Fe}\mathrm{As}\hspace{5.500 pt}$ & Total & & \multirow{2}{*}{High pressure}\\
-29.95 & \hspace{3.207 pt}-2.82\hspace{10.985 pt}\hspace{9.374 pt}-14.07\hspace{17.151 pt}\hspace{9.221 pt}-4.08\hspace{16.999 pt}\hspace{8.929 pt}-9.02\hspace{8.929 pt} & -30.00 & 0.05 & \\
\hline
$\mathrm{Tm}\mathrm{O}\mathrm{Fe}\mathrm{As}$ & $\hspace{5.500 pt}\frac{1}{3}\mathrm{Fe}\hspace{5.500 pt}$+$\hspace{5.500 pt}\frac{1}{3}\mathrm{Tm}_{2}\mathrm{O}_{3}\hspace{5.500 pt}$+$\hspace{5.500 pt}\frac{1}{3}\mathrm{Tm}\mathrm{As}\hspace{5.500 pt}$+$\hspace{5.500 pt}\frac{2}{3}\mathrm{Fe}\mathrm{As}\hspace{5.500 pt}$ & Total & &\multirow{2}{*}{High pressure} \\
-29.89 & \hspace{3.207 pt}-2.82\hspace{10.985 pt}\hspace{11.790 pt}-14.07\hspace{19.568 pt}\hspace{11.637 pt}-4.04\hspace{19.415 pt}\hspace{8.929 pt}-9.02\hspace{8.929 pt} & -29.96 & 0.07 & \\
\hline
$\mathrm{Yb}\mathrm{O}\mathrm{Fe}\mathrm{As}$ & $\hspace{5.500 pt}\mathrm{Yb}\mathrm{O}\hspace{5.500 pt}$+$\hspace{5.500 pt}\mathrm{Fe}\mathrm{As}\hspace{5.500 pt}$ & Total & & \multirow{2}{*}{no} \\
-25.13 & \hspace{2.861 pt}-13.06\hspace{10.639 pt}\hspace{3.236 pt}-13.54\hspace{3.236 pt} & -26.60 & 1.47 & \\
\hline
$\mathrm{Y}\mathrm{O}\mathrm{Fe}\mathrm{As}$ & $\hspace{5.500 pt}\frac{1}{3}\mathrm{Fe}\hspace{5.500 pt}$+$\hspace{5.500 pt}\frac{1}{3}\mathrm{Y}_{2}\mathrm{O}_{3}\hspace{5.500 pt}$+$\hspace{5.500 pt}\frac{1}{3}\mathrm{Y}\mathrm{As}\hspace{5.500 pt}$+$\hspace{5.500 pt}\frac{2}{3}\mathrm{Fe}\mathrm{As}\hspace{5.500 pt}$ & Total & & \multirow{2}{*}{no} \\
-31.71 & \hspace{3.207 pt}-2.82\hspace{10.985 pt}\hspace{7.887 pt}-15.19\hspace{15.665 pt}\hspace{7.735 pt}-4.74\hspace{15.512 pt}\hspace{8.929 pt}-9.02\hspace{8.929 pt} & -31.79 & 0.08 & \\
\hline
\hline   
\end{tabular}
\label{tn1}
\end{table*}

\section{Test and applications}

\begin{figure*}
% \flushright
\centering
%\begin{center}
\includegraphics [width=6.95in]{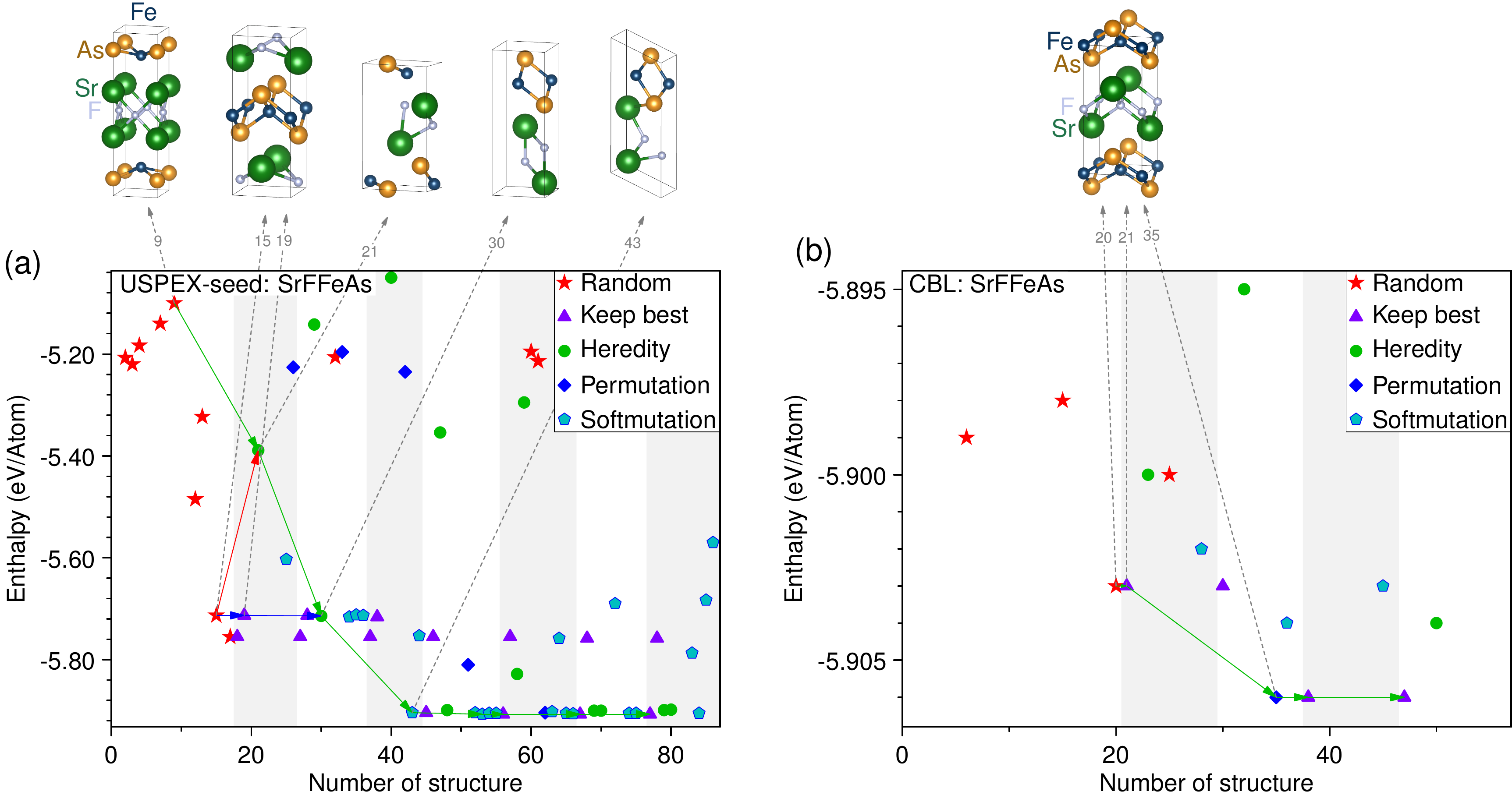}
\caption{The comparative test of SrFFeAs. The abscissa value of colored marks is the formation enthalpy of the corresponding structure. The colored arrow lines label the evolution route of the final best structure. The inherited structure contains genes from two different structures of the previous generation, such as the structures 21 and 30 in panel (a). The structures of different generations are background alternately by white and gray stripes. The two lowest enthalpy structures are kept as the best in the next generation. The random structures 4, 6, 9, 15, and 20 in panel (b) are already the target 1111 phase.}\label{f2}
%\end{center}
\end{figure*}

The structure prediction with the above model is realized based on the USPEX module of the source code of USPEX9.4 \cite{10.1063/1.2210932,10.1021/ar1001318}, which we call the ConfPred code for convenience. The VASP code is employed for structure relaxation. As tests, a series of existing iron-based superconductors are searched with the ConfPred code. The test conditions are: The fixed composition mode is adopted for all CSP calculations; The initial lattice is set to be tetragonal; Twenty random structures are generated in the first generation, and ten structures are produced in each subsequent generation by soft mutation (20\%), permutation (20\%), heredity (20\%), random (20\%), and keep best (20\%); The variable unit is embedded with the minimum ions distance between the fixed unit and the variable unit to be 1 \AA (As we verified, this value in range 0.5 to 2 is also workable.); The convergence condition of the iteration is that the best structure keeps unchanged for five generations; All the structures are in 1$\times$1 lattice; The lattice parameters, including the positions of all ions, are relaxed with the VASP code. 

Our tests include but are not limited to the following existing materials: LaOFeAs, Ba$_2$Ti$_2$Fe$_2$As$_4$O \cite{Sun.2012}, Sr$_3$Sc$_2$Fe$_2$As$_2$O$_5$ \cite{Zhu.2009} and Sr$_4$V$_2$Fe$_2$As$_2$O$_6$ \cite{Zhu.2009w08} in iron-based superconductors, and Sr$_2$Cr$_3$As$_2$O$_2$ \cite{10.1103/physrevb.92.205107}, Sr$_4$Sc$_2$Cr$_2$As$_2$O$_6$, Ba$_3$Sc$_2$Cr$_2$As$_2$O$_5$ in Cr-based layered materials \cite{Naik.2021}. The ConfPred code can easily find the target structures and nearly has a 100\% success rate in the repeated tests. The stability of inputs, such as the size of the confined space, the initial lattice parameters, and the distance between ions, are tested. We verified that this model is not sensitive to input parameters. For example, we have adjusted the lattice parameters a and c of the variable unit by up to $\pm$10\% and $\pm$30\% respectively. The ConfPred code can also find the target structures accurately. With such a strong fault tolerance, it’s hard to make a mistake here. In addition, the distance of ions can remain their default values. 
%Moreover, in most cases, only moderate accuracy is required for structural relaxation, of which the confined space may play a role.

\begin{figure*}
% \flushright
\centering
%\begin{center}
\includegraphics [width=6.95in]{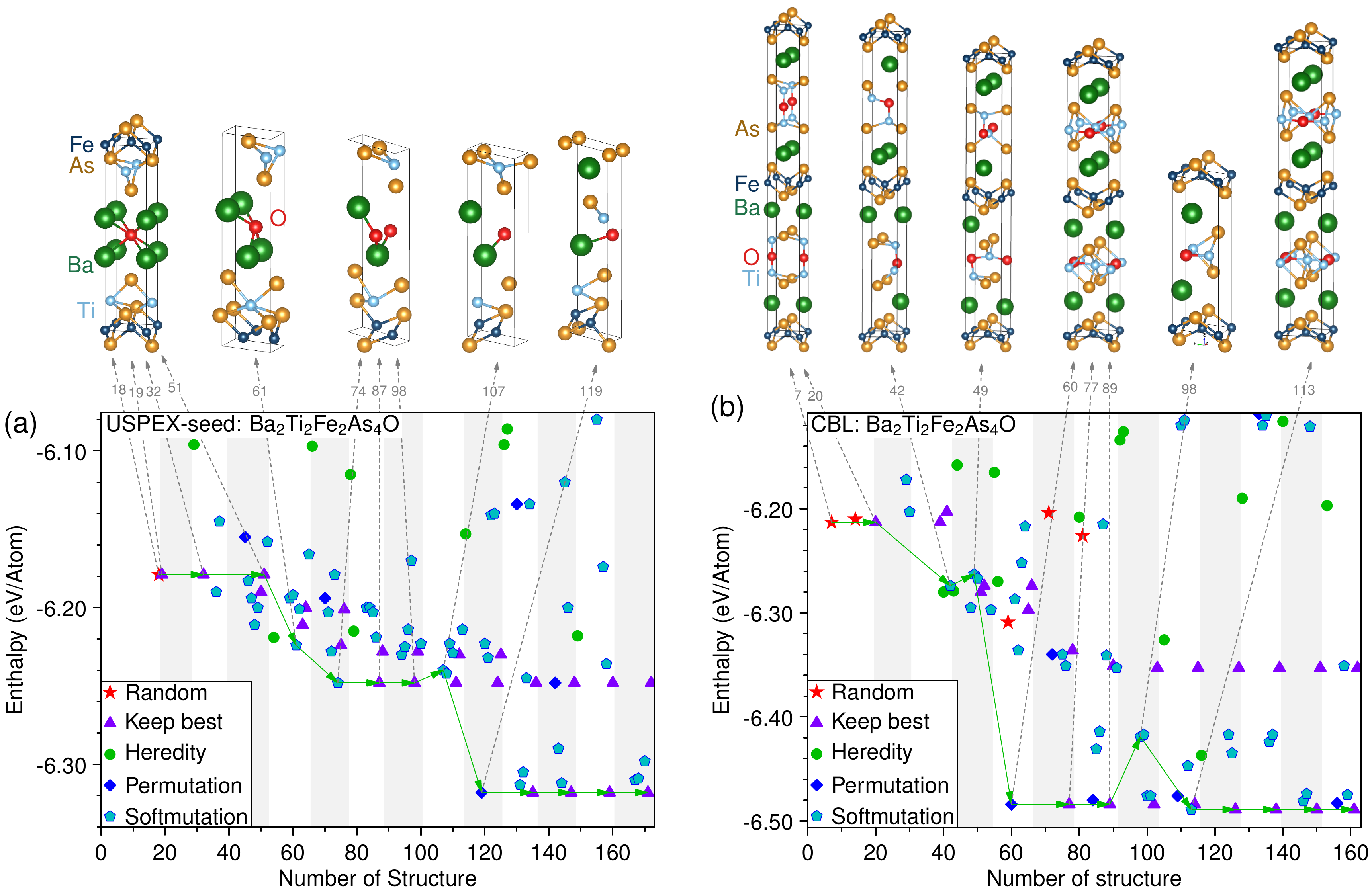}
\caption{The comparative test of Ba$_2$Ti$_2$Fe$_2$As$_4$O. (a) The evolution of the final best structure starts at a seed structure 18 that includes the Fe$_2$As$_2$ block layer. Please refer to the Appendix for the corresponding results of the test without seed. Please refer to Fig. \ref{f2} for other details about the data.}\label{f3}
%\end{center}
\end{figure*}

We also conduct a comparative test on SrFFeAs and Ba$_2$Ti$_2$Fe$_2$As$_4$O. One side is the ConfPred code, and the other is the original USPEX9.4 code. For the sake of fairness, the test of the original USPEX code is further divided into two parts, namely without seed and with a seed that includes the Fe$_2$As$_2$ block layer. The above test conditions remain unchanged here. The main results of the comparative test are listed in table \ref{t1}. For SrFFeAs, the seed we introduced is thrown out by the mechanism of heredity because it is substituted by a better random structure 15. For Ba$_2$Ti$_2$Fe$_2$As$_4$O, the seed is employed, and it reduces the structure enthalpy in the evolution process. The seed technology is helpful, but the effect is limited for searching large structures. Finally, only the ConfPred code finds the two target structures. 

Fig. \ref{f2} shows the prediction process of SrFFeAs for the different sides. 
As for the CSP of USPEX9.4 with seed in panel (a), the second generation (structure 21) along the evolution line is inherited from structures 9 and 15, and the third generation (structure 30) is inherited from structures 21 and 19. The structures 15 and 19 are almost the 1111 phase, and only the F$^-$ ions are not stated in a plane, which should be a metastable state. The structure 30 has already formed the Fe$_2$As$_2$ block layer and Sr$_2$F$_2$ layer. With further soft mutation, the target structure is obtained at the fifth generation in the form of reduced cell. In this CSP, the context of the metastable state is broken by genetic heredity, and the right phase is found by soft mutation. As for the CSP with the ConfPred code in panel (b), the target phase is always randomly generated at the first generation. 
As a whole, the target iron-based 1111 phase can be found for the two sides. The original USPEX code converged at the 8th generation. In contrast, the ConfPred code converged at the fifth generation, and the targeted structure can always be generated at the first generation. Usually, the enthalpy of the random structures generated by the ConfPred code is much lower than that of the original code because the fixed unit greatly reduces the size of the configuration space of ions.

\begin{table}[]
\caption{A search of ternary iron-based superconductors. The fixed structure unit is the Fe$_2$As$_2$ block layer for all of the structures except for LaFe$_3$As$_3$ with a fixed unit of La. We have also taken the cations (La, Ba, K, Th) as the fixed unit, but the structures found are either the same or higher in enthalpy except for LaFe$_3$As$_3$.}
\begin{tabular}{lccc}
\toprule
AFe$_m$As$_n$ & \begin{tabular}[c]{@{}c@{}} E$_\text{form}$\\  (eV/formula)\end{tabular} & Best structure & Space group \\
\midrule
LaFeAs$_2$         &  0.31    & 112 phase   &  129 (P4/nmm)    \\
BaFeAs$_2$         &  0.12    & 112 phase   &  129 (P4/nmm)    \\
BaFe$_2$As$_2$     & -0.79    & 122 phase   &  139 (I4/mmm)    \\
La$_2$Fe$_2$As$_3$ &  1.12    & Fig. 4(a)   &  139 (I4/mmm)    \\
LaFe$_3$As$_3$     &  0.98    & Fig. 4(b)   &   38 (Amm2)       \\
KFe$_4$As$_3$      &  0.75    & Fig. 4(c)   &  123 (P4/mmm)    \\
ThFe$_4$As$_4$     &  2.00    & Fig. 4(d)   &  129 (P4/nmm)    \\
BaFe$_5$As$_4$     &  2.50    & Fig. 4(e)   &   99 (P4mm)      \\   
\bottomrule    
\end{tabular}
\label{t2}
\end{table}

\begin{table*}
\caption{The symbol $^@$ label the already existing structures, and the others are the structures newly found by the ConfPred code. The functions to calculate the formation enthalpy E$_\text{form}$ are listed in table \ref{t4} of the Appendix.}
\begin{tabular}{lcccccl}
\toprule
Chemical formula  &\tabincell{c}{\quad E$_\text{form}$ \\ \quad (eV/formula) }& \quad a/b ($\mathring{A}$) & \quad c ($\mathring{A}$)   & \quad Space group  & Total generations & \quad Structure\\
\midrule
$^@$LaFeAs$_2$     &\quad  \,\,0.31    & \quad 3.99        & \quad 10.3 & \quad 129 (P4/nmm) & 10 & \quad 112 phase  \\
$^@$BaFeAs$_2$     &\quad  \,\,0.12    & \quad 3.98        & \quad 11.7 & \quad 129 (P4/nmm) & 10 & \quad 112 phase  \\
La$_2$Fe$_2$As$_3$ & \quad \,\,0.15    & \quad 3.99        & \quad 17.4 & \quad 139 (I4/mmm) & 5  & \quad Fig. 4(a)  \\
Eu$_2$Fe$_2$As$_3$ & \quad \,\,0.36    & \quad 3.98        & \quad 20.7 & \quad 139 (I4/mmm) & 5  & \quad Fig. 4(a)  \\
KFe$_4$As$_3$      &\quad  \,\,0.75    & \quad 3.79        & \quad 9.71 & \quad 123 (P4/mmm) & 5  & \quad Fig. 4(c)  \\
ThFe$_4$As$_4$     &\quad  \,\,2.43    & \quad 4.01        & \quad 16.2 & \quad 129 (P4/nmm) & 6  & \quad Fig. 4(d)  \\
La$_2$F$_2$Fe$_2$As$_3$ &\quad\,\,1.84 & \quad 3.87/4.05   & \quad 18.7 & \quad  38 (Amm2)   & 22 & \quad Fig. 4(f)  \\
La$_2$O$_2$ClFeAs  &\quad\,\,0.09      & \quad 4.04        & \quad 15.3 & \quad 129 (P4/nmm) & 8  & \quad Fig. 4(g)  \\
KLa$_2$O$_2$ClFe$_2$As$_2$& \quad\,\,1.35  & \quad 3.94    & \quad 12.3 & \quad  99 (P4mm)   & 11 & \quad Fig. 4(h)  \\
LiOFe$_2$As        &\quad  \,\,0.36    & \quad 3.68        & \quad 8.97 & \quad 129 (P4/nmm) & 11  & \quad Fig. 4(i)  \\
LiOMn$_2$As        &\quad  \,\,0.06    & \quad 3.65        & \quad 9.34 & \quad 129 (P4/nmm) & 11  & \quad Fig. 4(i)  \\
Li$_4$OFe$_2$As$_2$& \quad \,\,0.08    & \quad 3.82        & \quad 15.5 & \quad 139 (I4/mmm) & 10  & \quad Fig. 4(j)  \\
\bottomrule
\end{tabular}
\label{t3}
\end{table*}

Fig. \ref{f3} shows the prediction process of Ba$_2$Ti$_2$Fe$_2$As$_4$O \cite{Sun.2012} for two sides. 
As for panel (a), the seed structure 18 is the start point of the optimal evolution line and has been kept best for three generations. By soft mutation, a more stable structure is produced at the fifth generation, in which the bonding angle is optimized. It falls into another metastable state and has been trapped for four generations. By the permutation of Ba and Ti, it successfully strides over the trap at the 10th generation. As limited by the computation scale, this metastable state is kept as the final convergent state.
As for panel (b), the evolution line starts at a good-quality structure that the Ba ions are at the right sites, and the Ti, O are bonded together. It is worth mentioning that this instance is almost the worst one in the repeated tests. In most cases, some much better random structures can be generated at the first generation, which can be ascribed to the configurational reduction by the fixed structure unit. By the soft mutation in the third generation, a better Ti-O bonding is formed. A gorgeous evolution occurs at the fifth generation, which successfully forms the Ti2O plane. The target phase is formed at the fifth generation. It is worth noting that structure 98 is the 22241 phase in Niggli reduced cell. 
As a whole, the original USPEX code reached convergence after 14 generation, but didn't converge to the target phase Ba$_2$Ti$_2$Fe$_2$As$_4$O \cite{Sun.2012}. Under this calculation scale, the original code has a meager success rate in finding the target Ba$_2$Ti$_2$Fe$_2$As$_4$O phase. The ConfPred code found the target structure Ba$_2$Ti$_2$Fe$_2$As$_4$O at the fifth generation and reached convergence at the 13th generation. As revealed by the repeated tests, the ConfPred code is very stable and insensitive to the input parameters. The structure Ba$_2$Ti$_2$Fe$_2$As$_4$O can be decomposed into BaTi$_2$As$_2$O$+$BaFe$_2$As$_2$, so we searched BaTi$_2$As$_2$O with the original code. It failed many times under the calculation scale we applied. However, searching for BaTi$_2$As$_2$O phase between the Fe$_2$As$_2$ layers according to the ConfPred code is fast and reliable, indicating that the two-dimensional confined space plays an important role in the arrangement of ions during the relaxation process, and significantly reduces the number of metastable states. In addition, the original USPEX code failed again to find the 22241 phase with increasing the calculation scale by two times. The ConfPred code can also find the 22241 phase by reducing the calculation scale two times. Undoubtedly, if the calculation scale is increased further, the success rate of the original USPEX code in searching the above structures would improve. The above comparative tests are based on the same input parameters and calculation scale, so the results are of practical significance and reference value.

\begin{figure*}
\centering
%\begin{center}
\includegraphics [width=6.5in]{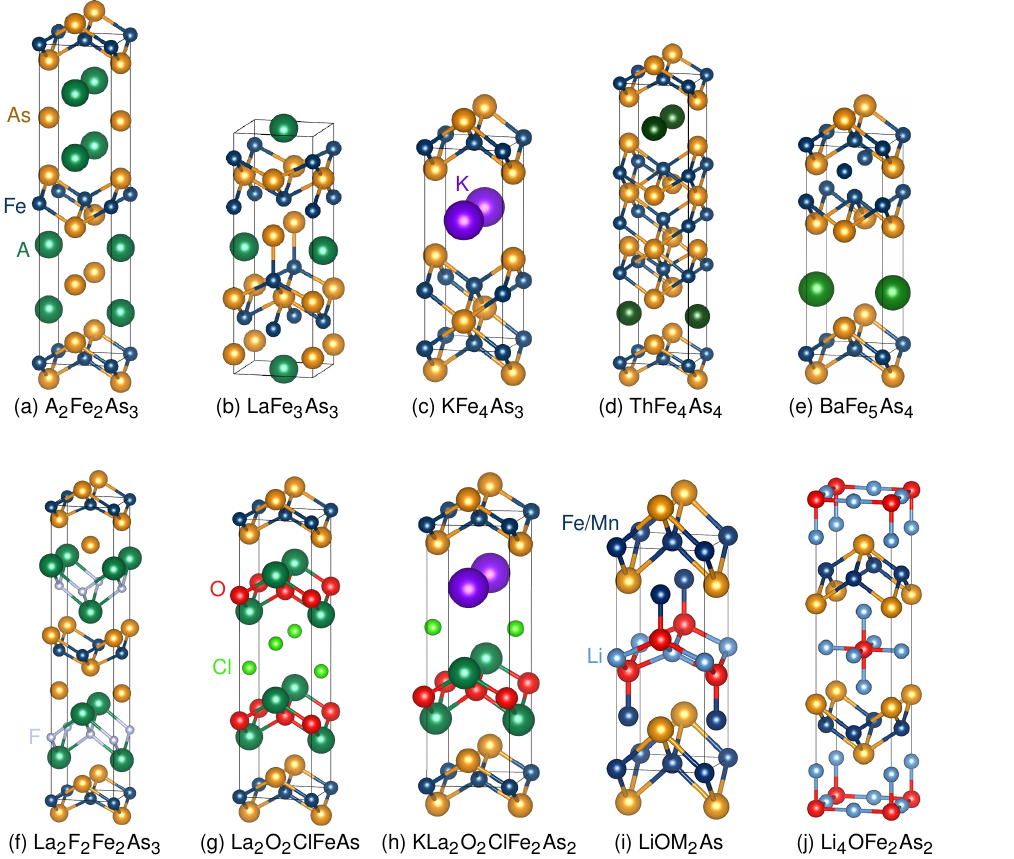}
\caption{The picture of the structures in tables \ref{t2} and \ref{t3}. (a) ``A'' refers to La or Ba. The fixed structure unit is the Fe$_2$As$_2$ block layer for all the structures except for LaFe$_3$As$_3$ with a fixed unit of La. (i) M refers to Fe or Mn.}
\label{f4}
%\end{center}
\end{figure*}

We launched a practical prediction of new superconductors with the ConfPred code. We first searched the multi-composition ternary iron-based superconductors. The already existing 112 phase and 122 phase are found successfully. Interestingly, a newly 223 structure has a much lower formation enthalpy than 112 phase, which will be discussed in the following. As the proportion of Fe and As increases, the lattice formation energy decreases firstly and then increases continuously. The lowest formation enthalpy occurs at the proportion 2:2.
Besides, we have also done some other searches. Some of the results are listed in table \ref{t3}. 
By embedding some cations (A$^{n+}$, A refers to La, Ba, K, Th) and As anions between Fe$_2$As$_2$ layers, we finally found the 112, 122 phase of the iron-based superconductors \cite{Vinod.2015,Kang.2017,10.1103/physrevlett.101.057006}, and some other structures, such as La$_2$Fe$_2$As$_3$, LaFe$_3$As$_3$, K$_2$Fe$_4$As$_3$, Th$_2$Fe$_4$As$_4$ and Ba$_2$Fe$_5$As$_4$ (Please refer to Fig. \ref{f4} for their pictures). Actually, we have designed the 223 and 143 configuration of structure years ago \cite{Jiang.2013i1j}. The formation enthalpy of LaFeAs$_2$ and BaFeAs$_2$ is about 0.3 eV, so these two structures are unstable. Experimentally, the parent phase has not been synthesized by routine method, while the 112 phase can be formed by doping \cite{Yakita.2014,10.1063/1.4941277}. Interestingly, the formation enthalpy of La$_2$Fe$_2$As$_3$ is much lower than the 112 phase, suggesting that it potentially can be synthesized by doping or pressing as well. As for the 223 configuration, another potential structure Eu$_2$Fe$_2$As$_3$ deserves scrutiny. Meanwhile, more extensive element substitution is worth considering. Although the formation enthalpy of LaFe$_3$As$_3$, K$_2$Fe$_4$As$_3$, Th$_2$Fe$_4$As$_4$ and Ba$_2$Fe$_5$As$_4$ are too high to be synthesized in normal ways, there are also bright spots. Comparing to the standard Fe$_2$As$_2$ block layer, the positions of As and Fe are swapped in the structure LaFe$_3$As$_3$. In the structure K$_2$Fe$_4$As$_3$, two layers of Fe$_2$As$_2$ are combined together by sharing an As ion. In Th$_2$Fe$_4$As$_4$, more layers of Fe$_2$As$_2$ are combined.

For another recipe, we got the structures LaFFe$_2$As$_2$, LaF$_2$Fe$_2$As$_2$ and La$_2$F$_2$Fe$_2$As$_3$. Their formation enthalpies are too big to arouse the interest of experimenters. Their potential value may be found in FeSe superconductors.  

Another metastable structure La$_2$O$_2$ClFeAs with five different elements were found by embedding La$^{3+}$, O$^{2-}$ and Cl$^-$ ions between the Fe$_2$As$_2$ layer, whose formation enthalpy is 0.09 eV. The variable unit in this structure is the existing material LaOCl. This configuration of structure deserves further exploration theoretically and experimentally.

A similar strategy found some Li-based metastable structures. They are LiOFe$_2$As, LiOMn$_2$As and Li$_4$OFe$_2$As$_2$. They are highly possible to be synthesized by doping or high pressure.

\section{Discussion}
Obviously, the model with two-dimensional confined space has significant advantages in predicting the structure of layered 3D materials. Even the complex iron-based superconductor can be easily found by this model of CSP, making it has a promising application prospect. The model is specifically designed for structure prediction of layered superconductors. The fixed unit dramatically reduces the size of the configuration space of ions, which significantly brings down the difficulty of the search problem. In addition, the self-assembly behavior in confined space guides the ions to arrange themselves in order, which is helpful for the ranking problem. If the fixed structure unit is a non-superconducting block layer, this model can also search for a specific type of new superconducting layer and other layered materials.

\section{Summary}
We proposed a model to accelerate the prediction of layered high-temperature superconductors utilizing confinement self-assembly. We carried it out based on the USPEX9.4 code. The tests indicate that this model can significantly reduce the machine time and has a very high success rate even at a tiny computation scale. Based on it, the prediction of the layered high-temperature superconductors of up to six elements can be easily realized almost on a single machine. With the ConfPred code, we have found several potential new superconducting structures at atmospheric pressure so far. We share the modified code in the author’s website.

%We acknowledge the financial support from the .

 \bibliographystyle{apsrev4-2}
%  %\bibliographystyle{urstr}
 \bibliography{ConfPred}
\section{Appendix}

\begin{figure*}
\centering
%\begin{center}
\includegraphics [width=6.95in]{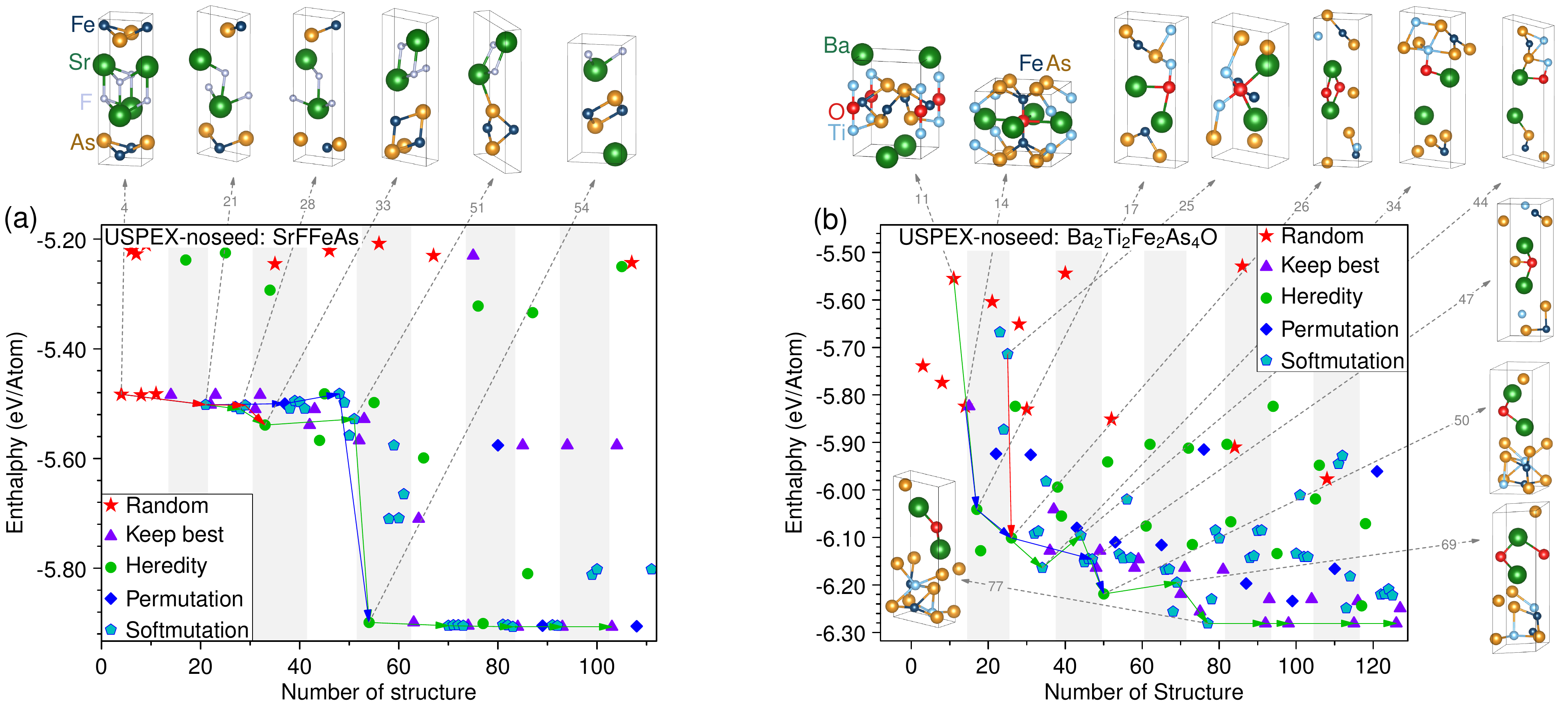}
\caption{The prediction process of SrFFeAs and Ba$_2$Ti$_2$Fe$_2$As$_4$O without seed.}
\label{f5}
%\end{center}
\end{figure*}

\begin{table*}[]
\caption{This table lists the functions that we used to calculate the formation enthalpy of the structures. The most stable compounds are selected as decomposers based on our experience. If the formation enthalpy we calculated with a certain function is minus or close to zero, then we will try our best to find a better function to maximize the formation enthalpy.}
\begin{tabular}{c|cc|c}
\hline
\hline
Compound \&      & Decomposers \&    &   &  E$_\text{form}$  \\
energy (eV/formula) & energy (eV/formula)& & (eV/formula)\\
\hline   
$\mathrm{La}\mathrm{Fe}\mathrm{As}_{2}$ & $\hspace{5.500 pt}\mathrm{La}\mathrm{As}\hspace{5.500 pt}$+$\hspace{5.500 pt}\mathrm{Fe}\mathrm{As}\hspace{5.500 pt}$ & Total & \\
-25.95 & \hspace{3.792 pt}-12.72\hspace{11.569 pt}\hspace{3.236 pt}-13.54\hspace{3.236 pt} & -26.26 & 0.31 \\
\hline
$\mathrm{Ba}\mathrm{Fe}\mathrm{As}_{2}$ & $\hspace{5.500 pt}\frac{1}{2}\mathrm{Ba}\mathrm{As}_{2}\hspace{5.500 pt}$+$\hspace{5.500 pt}\frac{1}{2}\mathrm{Ba}\mathrm{Fe}_{2}\mathrm{As}_{2}\hspace{5.500 pt}$ & Total & \\
-21.95 & \hspace{11.451 pt}-6.80\hspace{19.229 pt}\hspace{16.957 pt}-15.27\hspace{16.957 pt} & -22.07 & 0.12 \\
\hline
$\mathrm{La}_{2}\mathrm{Fe}_{2}\mathrm{As}_{3}$ & $\hspace{5.500 pt}\mathrm{Fe}\hspace{5.500 pt}$+$\hspace{5.500 pt}2\mathrm{La}\mathrm{As}\hspace{5.500 pt}$+$\hspace{5.500 pt}\mathrm{Fe}\mathrm{As}\hspace{5.500 pt}$ & Total & \\
-46.34 & \hspace{0.014 pt}-8.47\hspace{7.792 pt}\hspace{6.292 pt}-25.45\hspace{14.069 pt}\hspace{3.236 pt}-13.54\hspace{3.236 pt} & -47.45 & 1.12 \\
\hline
$\mathrm{K}\mathrm{Fe}_{4}\mathrm{As}_{3}$ & $\hspace{5.500 pt}\frac{3}{2}\mathrm{Fe}\hspace{5.500 pt}$+$\hspace{5.500 pt}\frac{1}{2}\mathrm{Fe}\mathrm{As}_{2}\hspace{5.500 pt}$+$\hspace{5.500 pt}\mathrm{K}\mathrm{Fe}_{2}\mathrm{As}_{2}\hspace{5.500 pt}$ & Total & \\
-50.05 & \hspace{0.707 pt}-12.70\hspace{8.485 pt}\hspace{11.172 pt}-9.29\hspace{18.950 pt}\hspace{11.611 pt}-28.81\hspace{11.611 pt} & -50.81 & 0.75 \\
\hline
$\mathrm{Th}\mathrm{Fe}_{4}\mathrm{As}_{4}$ & $\hspace{5.500 pt}\frac{4}{3}\mathrm{Fe}\hspace{5.500 pt}$+$\hspace{5.500 pt}\frac{1}{3}\mathrm{Th}_{3}\mathrm{As}_{4}\hspace{5.500 pt}$+$\hspace{5.500 pt}\frac{8}{3}\mathrm{Fe}\mathrm{As}\hspace{5.500 pt}$ & Total & \\
-62.15 & \hspace{0.707 pt}-11.29\hspace{8.485 pt}\hspace{12.235 pt}-16.75\hspace{20.013 pt}\hspace{6.429 pt}-36.10\hspace{6.429 pt} & -64.14 & 2.00 \\
\hline
$\mathrm{La}\mathrm{F}\mathrm{Fe}_{2}\mathrm{As}_{2}$ & $\hspace{5.500 pt}\frac{2}{3}\mathrm{Fe}\hspace{5.500 pt}$+$\hspace{5.500 pt}\frac{1}{3}\mathrm{La}\mathrm{F}_{3}\hspace{5.500 pt}$+$\hspace{5.500 pt}\frac{2}{3}\mathrm{La}\mathrm{As}\hspace{5.500 pt}$+$\hspace{5.500 pt}\frac{4}{3}\mathrm{Fe}\mathrm{As}\hspace{5.500 pt}$ & Total & \\
-37.92 & \hspace{3.207 pt}-5.65\hspace{10.985 pt}\hspace{9.269 pt}-9.08\hspace{17.047 pt}\hspace{9.485 pt}-8.48\hspace{17.263 pt}\hspace{6.429 pt}-18.05\hspace{6.429 pt} & -41.26 & 3.34 \\
\hline
$\mathrm{La}\mathrm{F}_{2}\mathrm{Fe}_{2}\mathrm{As}_{2}$ & $\hspace{5.500 pt}\frac{1}{3}\mathrm{Fe}\hspace{5.500 pt}$+$\hspace{5.500 pt}\frac{2}{3}\mathrm{La}\mathrm{F}_{3}\hspace{5.500 pt}$+$\hspace{5.500 pt}\frac{1}{3}\mathrm{La}\mathrm{As}\hspace{5.500 pt}$+$\hspace{5.500 pt}\frac{5}{3}\mathrm{Fe}\mathrm{As}\hspace{5.500 pt}$ & Total & \\
-46.27 & \hspace{3.207 pt}-2.82\hspace{10.985 pt}\hspace{6.769 pt}-18.17\hspace{14.547 pt}\hspace{9.485 pt}-4.24\hspace{17.263 pt}\hspace{6.429 pt}-22.56\hspace{6.429 pt} & -47.80 & 1.53 \\
\hline
$\mathrm{La}_{2}\mathrm{F}_{2}\mathrm{Fe}_{2}\mathrm{As}_{3}$ & $\hspace{5.500 pt}\frac{1}{3}\mathrm{Fe}\hspace{5.500 pt}$+$\hspace{5.500 pt}\frac{2}{3}\mathrm{La}\mathrm{F}_{3}\hspace{5.500 pt}$+$\hspace{5.500 pt}\frac{4}{3}\mathrm{La}\mathrm{As}\hspace{5.500 pt}$+$\hspace{5.500 pt}\frac{5}{3}\mathrm{Fe}\mathrm{As}\hspace{5.500 pt}$ & Total & \\
-58.68 & \hspace{3.207 pt}-2.82\hspace{10.985 pt}\hspace{6.769 pt}-18.17\hspace{14.547 pt}\hspace{6.985 pt}-16.96\hspace{14.763 pt}\hspace{6.429 pt}-22.56\hspace{6.429 pt} & -60.52 & 1.84 \\
\hline
$\mathrm{La}_{2}\mathrm{O}_{2}\mathrm{Cl}\mathrm{Fe}\mathrm{As}$ & $\hspace{5.500 pt}\mathrm{La}\mathrm{Cl}\mathrm{O}\hspace{5.500 pt}$+$\hspace{5.500 pt}\mathrm{La}\mathrm{Fe}\mathrm{As}\mathrm{O}\hspace{5.500 pt}$ & Total & \\
-51.45 & \hspace{6.958 pt}-21.25\hspace{14.736 pt}\hspace{12.750 pt}-30.29\hspace{12.750 pt} & -51.54 & 0.09 \\
\hline
\hline   
\end{tabular}
\label{t4}
\end{table*}

Fig. \ref{f5} shows the prediction process of SrFFeAs and Ba$_2$Ti$_2$Fe$_2$As$_4$O without seed. As for Fig. \ref{f5} (a), it found the target 1111 phase at the sixth generation by heredity. As for Fig. \ref{f5} (b), It has been run for 11 generations. It did not find the target 22241 phase and failed to form the Fe$_2$As$_2$ block layer. Too many metastable states should be the main reason behind it. From the evolution line shown, the prediction of Ba$_2$Ti$_2$Fe$_2$As$_4$O without other technology is run out of the ability of the current CSP method, verifying that the prediction of quaternary structures is too much of a challenge to achieve by the general ways. The model we proposed would change this situation in the field of predicting intergrowth materials. Table \ref{f4} shown the ways to calculate the formation enthalpies. The choice of the decomposers is critical for the accuracy of the formation enthalpy. To get the reliable values, we have tried multiple combinations of different decomposers to eliminate the possible mischoose as much as possible.

The lattice relaxation ware divided into five steps for each phase. All of the tests were conducted on a server machine with 20 cores.

\end{document}